# Moving Past the Minimum Information Partition: How To Quickly and Accurately Calculate Integrated Information

**Daniel Toker and Friedrich Sommer**
Helen Wills Neuroscience Institute
University of California, Berkeley
danieltoker@berkeley.edu

## Abstract

An outstanding challenge with the Integrated Information Theory of Consciousness (IIT) is to find a way of rapidly and accurately calculating integrated information from neural data. A number of measures of integrated information based on time series data have been proposed, but most measures require finding the "Minimum Information Partition" of a network, which is computationally expensive and not practical for real brain data. Here, we introduce a novel partition, the Maximum Modularity Partition, across which to quickly calculate integrated information. We also introduce a novel detection task on simulated data to evaluate the performance of integrated information measures across different partitions. We show that integrated information can be reliably and quickly calculated across the Maximum Modularity Partition, as well as the previously proposed atomic partition, even in relatively large networks, which constitutes an advance in researchers' ability to empirically test the predictions of IIT in real brain data.

## 1 Introduction

Giulio Tononi's Integrated Information Theory of consciousness (IIT) assumes that the thalamocortical system can generate conscious experience because of the way it processes information. Specifically, the theory claims that the thalamocortical system integrates information as a system over and above its parts, and that the amount of integrated information in a brain's thalamocortical system (or some significant part of it) should track that brain's level of consciousness [1], [2], [3], [4]. Despite the attention that IIT has garnered from a wide variety of disciplines (e.g. [5] and [6]), little has been done to test the validity of the theory in real brain data.

Tononi's original measure of integrated information [1] has been impossible to apply to real brain data because it requires systematic perturbation of every node in a network. Tononi's more recent framework [4] is also practically infeasible for real brains, because it requires knowing the transition probability matrix of an entire neural network, which is impossible to determine in real neural systems.

These limitations have inspired several alternative measures of integrated information, which can in principle be computed from time series data [7], [8], [9], [10]. But these measures are also limited. While they do not require systematic perturbation of a system or knowledge of complete network functional architecture − and are therefore an improvement on Tononi's measures in terms of real-world applicability − they are still prohibitively slow to compute in actual brains. This is because, as with Tononi's measures, each of these alternative measures requires finding the "minimum information partition" (MIP) of a system.

The MIP is the network partition across which there is the least information flow. One of the intuitions of IIT is that information flow across the MIP should track how much information a system integrates or has the capacity to integrate, because it tells you how much information a system generates over and above how much information is generated by its subsystems, which are delineated by the most informative decomposition (i.e. the MIP) of the system.

There are two major problems with using the MIP as the partition across which to evaluate integrated information: 1) The computation time required for finding the MIP grows super-exponentially with system size [11], such that it is infeasible to find the MIP for anything but very, very small artificial networks, and 2) Conceivably, the partition across which there is the least information flow might be a single node isolated from the remainder of the system, which likely does not reflect how information is actually integrated in the brain. One alternative to finding the MIP is to find the minimum information *bi*partition (MIB) of a system, which is the bipartition across which there is the least information flow in a network (Figure 1B). If the MIB splits a system into two equal-sized halves, then, unlike the MIP, it does not run the risk of isolating one or a few nodes from the remainder of a system. The MIB is also faster to compute than the MIP, although the time it takes to find the MIB also grows exponentially with larger systems. As such, finding the MIB is impractical for real brains. Although Tegmark [11] has recently proposed a graph-theoretic approach to quickly approximate the MIB based on network structure, there are still conceptual issues with computing integrated information across the MIB: it is not clear *a priori* whether the MIB is a meaningful way to decompose a neural network because it is unlikely that functional subnetworks divide the brain exactly in half.

We propose an alternative network partition acrosss which to calculate integrated information, which is based on the graph-theoretic notion of modularity and bypasses the computational expense of finding the MIP and the artificiality of using the MIB. The modularity of a network quantifies the extent to which that network can be broken up into relatively discrete ”modules,” which are defined as densely interconnected communities of nodes. Many real-world networks have been found to follow a modular architecture, including yeast transcriptional networks [12] and metabolic networks [13]. The human brain also follows a broadly modular organization in its anatomy [14], [15], its patterns of gene expression [16], [17], [18], and its functional architecture [15], [19], [20], [21]. In particular, functional modules in the human brain are thought to represent relatively isolated subnetworks across which information is integrated [22], [21]. Decomposing a neural network into its ”Maximum Modularity Partition” (MMP) (Figure 1C) is therefore more congruent with what is known about information flow in the brain than is calculating integrated information across the MIB, because it is reflects the underlying functional architecture of neural networks. The MMP has the further advantage over the MIP and the MIB in being extremely quick to find [23], [24], [25], [26], [27].

Another fast alternative for partitioning networks is the ”atomic” partition, which has been proposed previously [10]. The atomic partition treats every node in a network as its own part; in other words, it is the complete partition of a network, such that no node is clustered together with any other node in a community or module (Figure 1D). While using the atomic partition is a fast alternative to using the MIP or MIB, local information processing in the brain is likely to be instantiated in communities of nodes rather than in single nodes (depending of course on what is being considered a ”node”), but we leave the atomic partition as an open option for fast empirical measurements of integrated information.

Although evaluating integrated information across the MMP or the atomic partition is promising in terms of practicality, it remains unclear in practice whether using these alternative partitions yield results as, more, or less accurate than calculating integrated information across the MIB, which, thus far, has been the primary alternative to the MIP [1], [7], [8], [9], [10], [28], [29], [30].

In this study, we compare three alternative network partitions: the MIB, the MMP, and the atomic partition. To do so, we introduce a novel detection task to address the practical applicability of computing integrated information across different partitions. For reference, we also test a measure of network integration that is not evaluated across any partition. Our task simply tests whether a given measure across a given partition can identify when a network has been structurally bipartitioned. Any good measure of integrated information, across whatever partition, should at least pass this test.

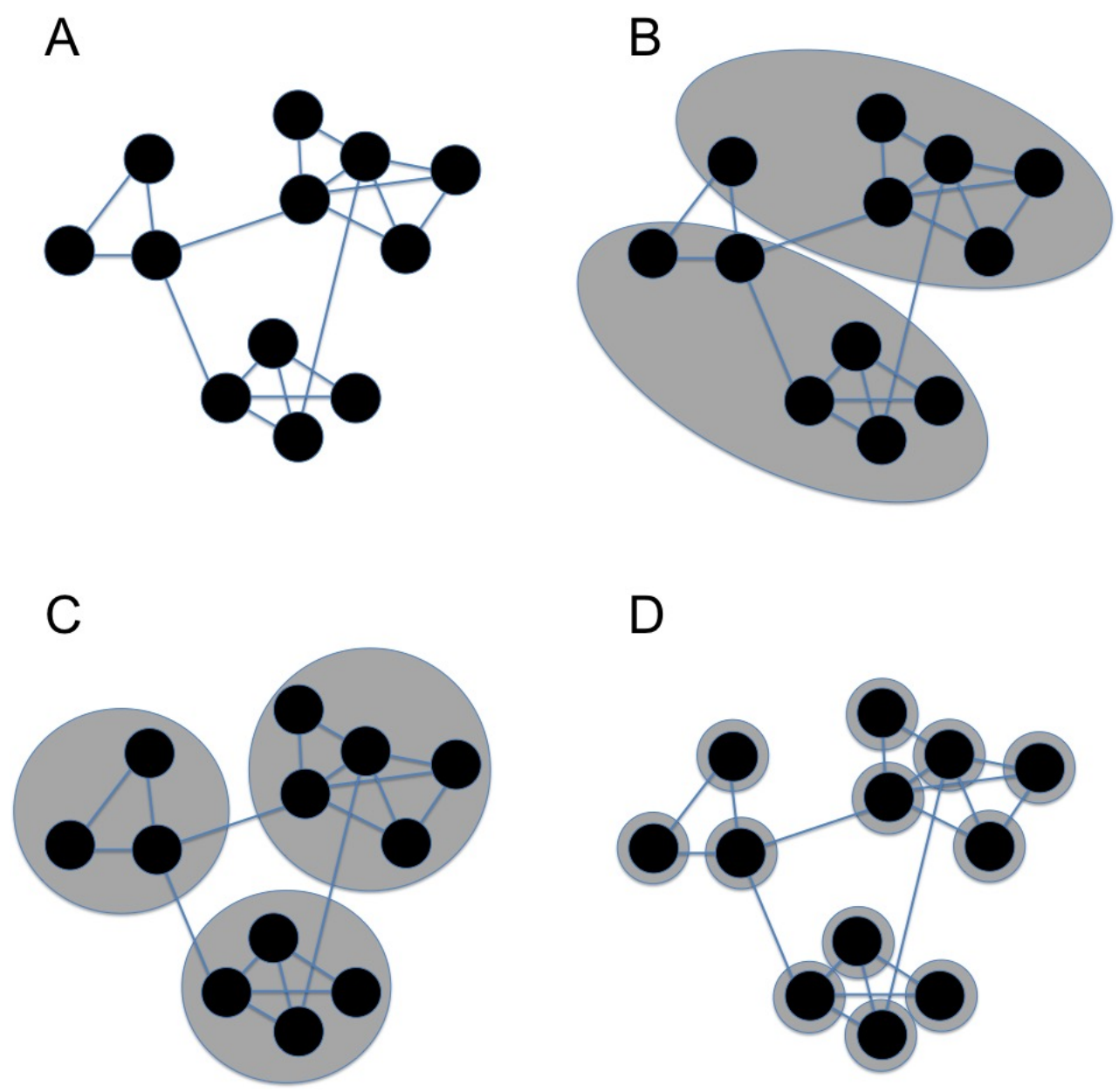

**Figure 1:** Network partition options. Gray ovals delineate the partition across which integrated information is evaluated. A) An example of a small network. In order to calculate the integrated information in this network, most measures require splitting the network into some partition and seeing how much more information the whole network carries than the sum of its parts. B) The "minimum information bipartition" (MIB) of the network. The MIB is the bipartition across which there is the least information flow. Finding the MIB requires trying every possible bipartition of a network, a process whose computation time increases exponentially with bigger networks, and which may not accurately reflect how information actually flows in the brain. C) The "Maximum Modularity Partition" (MMP). The MMP splits a network into modules. The definition of a module is that its nodes are more densely connected to each other than they are to nodes outside the module. The brain follows a broadly modular architecture in its anatomy, function, and patterns of gene expression, and so the MMP has the advantage over other partitions in being congruent with what is known about large-scale information flow in the brain. The MMP can also be quickly computed using any of a number of highly optimized algorithms. D) The "atomic partition." The atomic partition treats every node as its own part.

## 2 Methods

### 2.1 Measures of Integrated Information

There exist a variety of proposed measures of integrated information from time series data. For a thorough review of the current "landscape" of integrated information measures, see [11]. While some measures seem more attractive than others for *a priori* reasons, to our knowledge there have not yet been empirical tests of those measures' quality. Here, we test three previously proposed measures of integrated information from time series data: Causal Density, which is not strictly speaking a measure of integrated information but which has been shown to behave similarly to measures of integrated information [7], [8], $\Phi^*$ [10], and $\Phi^{\mathrm{AR}}$ [9]. While we could have chosen any of a variety of previously proposed measures, we picked these three because they are representative of a wide range of assumptions and approaches, as explained in greater detail below. In this section, we outline the basics of each of these three measures. For a more thorough mathematical explanation of the measures, see the Supplementary Materials S1.1.

#### Measure of Integrated Information: $\Phi^{\mathbf{AR}}$

Barrett and Seth [9] have proposed several measures of integrated information based on time series data. One of those measures, $\Phi^{\mathrm{AR}}$, is particularly attractive because, unlike other information-theoretic measures, it does not require obtaining empirical covariance matrices and can therefore easily be applied to non-Gaussian data. $\Phi^{\mathrm{AR}}$ is therefore worth investigating because neural time series often do not follow a Gaussian distribution: spike data tend to be Poisson-distributed [31], and both EEG [32] and BOLD fMRI data [33], [34] tend to be non-Gaussian.

$\Phi^{\mathrm{AR}}$ uses simple linear regression to estimate how much more predictive the present of a system is of its past as compared to how predictive its parts are of their own past. An intuitive way to conceptualize $\Phi^{\mathrm{AR}}$ is as follows:

$$\Phi^{\mathrm{AR}}_{\mathrm{partition}}(X) = \mathrm{Info}(X) - \mathrm{Info}(X_{\mathrm{partition}}), \tag{1}$$

,

where ”partition” refers to how the system is being decomposed into parts, and ”Info” is a non-Shannon measure of how informative a system is about itself over time. In other words, $\Phi^{\mathrm{AR}}$ quantifies the difference between how informative a system is about itself and the sum of how informative its parts are about themselves. See the Supplementary Materials S1.1 or Barrett and Seth [9] for a more formal explanation of $\Phi^{\mathrm{AR}}$.

Under Gaussian conditions, $\Phi^{\mathrm{AR}}$ is interpretable in terms of traditional Shannon information because it is based on differential entropy. Unfortunately, differential entropy can be negative. The consequence, as pointed out in [10] and [11], is that under Gaussian conditions, $\Phi^{\mathrm{AR}}$ can be negative. This is problematic, because integrated information cannot be negative [10], [11]. Under non-Gaussian conditions, $\Phi^{\mathrm{AR}}$ can still be negative, and its terms also deviate from classical differential entropy. Therefore, under non-Gaussian conditions, $\Phi^{\mathrm{AR}}$ and is no longer interpretable in terms of traditional Shannon information. That said, Barrett and Seth [9] convincingly argue that even under non-Gaussian conditions $\Phi^{\mathrm{AR}}$ is a meaningful measure of integrated information, because it still quantifies how informative a system is about itself relative to its parts.

### Measure of Integrated Information: $\Phi^*$

Oizumi et al [10] propose a novel measure of integrated information which is upper bounded by the maximum information in a system and lower bounded by zero, and which therefore does not violate the assumptions that are violated by $\Phi^{\mathrm{AR}}$. This new measure, $\Phi^*$, is based on the information-theoretic notion of mismatched decoding [35]:

$$\Phi^*_{\mathrm{partition}}(X^{t-\tau}; X^{\tau}) = I(X^{t-\tau}; X^{\tau}) - I^*_{\mathrm{partition}}(X^{t-\tau}; X^{\tau}) \tag{2}$$

where $t$ is the present of the system, $t - \tau$ is the past of the system, $I$ is the mutual information between the past and present of the intact system, and $I^*_{\mathrm{partition}}$ is the ”hypothetical” mutual information for mismatched decoding between the past and present of the system. For more details on calculating $I$ and $I^*_{\mathrm{partition}}$, see the Supplementary Materials S1.1.

$\Phi^*$ tracks the degree to which the past of a whole system is predictive of its present over and above the degree to which the past of its parts are predictive of their own present. Therefore, like $\Phi^{\mathrm{AR}}$, $\Phi^*$ is a way of measuring how much more information there is in the whole of a system than there is in the sum of its parts, except that the temporal directionality of the information is reversed.

Because $\Phi^*$ is lower-bounded by 0, it is more theoretically sound than $\Phi^{\mathrm{AR}}$ as a measure of integrated information. That said, the analytic computation of $\Phi^*$ requires that the data under consideration be Gaussian, which is an important limitation to consider when applying $\Phi^*$ to real neural data.

### Causal Density

Seth et al [7] propose a measure of ”causal density” (CD), which tracks the average strength of pairwise causal interactions in a network. Causal density leverages an econometric measure known as Granger causality $F$ [36], which measures the extent to which the past of a variable $X$ predicts the future of variable $Y$ over and above the degree to which the past of $Y$ predicts its own future. In this sense, Granger causality is an attempt to track the extent to which $X$ is predictively causal of $Y$.

Seth et al [7] define CD as the average pairwise Granger causality $F$ between every node in an $n$-node network:

$$CD(X) =: \frac{1}{n(n-1)} \sum_{i \neq j} F_{X_i \rightarrow X_j | X_{[ij]}} \tag{3}$$

(For details on how Granger causality $F$ is calculated, see the Supplementary Materials S1.1). Although CD is not in the strict sense a measure of integrated information, it behaves much like integrated information measures in the sense that it is low when a network is either highly integrated (meaning it has little information) or when a network is highly differentiated (meaning it has little integration) [7]. CD is compared to $\Phi^{\mathrm{AR}}$ and $\Phi^*$ because, unlike these measures, it is not evaluated across a network partition. That said, like many proposed measures, CD assumes that the underlying time series data is Gaussian.

### 2.2 The MMP: A Novel Partition for Calculating Integrated Information

As mentioned in the introduction, thus far the most common partition across which to calculate integrated information has been the minimum information bipartition (MIB) (Figure 2B) [1], [7], [8], [9], [10], [28], [29], [30]. As others have pointed out, the MIB can be prohibitively slow to compute, and so a previous attempt to speed up the calculation of integrated information focused on quickly identifying the MIB [11]. Little attention has been paid to alternative partitions, with the exception of Oizumi et al [10], in which the atomic partition was used a fast alternative to the MIB (Figure 2D).

Here, we introduce the maximum modularity partition (MMP) as a novel partition across which to calculate integrated information (Figure 2C). While the MMP has been studied extensively in network neuroscience (e.g. [15], [19], [20], [21]), it has not yet been studied in the context of Integrated Information Theory. We propose that calculating integrated information across the most modular decomposition of a neural network is a highly promising approach, in light of previous work that has shown information to flow across relatively discrete modules in the brain [22], [21].

There are a number of ways of finding the MMP of a network. We utilized the Louvain Method for detecting communities in graphs, which is a "greedy optimization" method and which seems to run on O(nlogn) time [25].

To find the maximum modularity partition (MMP), we first obtain an adjacency matrix from time series data. Specifically, we take the autocorrelation of the time series and binarize the resulting correlation matrix at a given threshold [37]. In this study, this threshold was set at statistical significance at an alpha of 0.05 and being in the upper half of the distribution of the absolute values of correlation coefficients. The resulting binary adjacency matrix is thought to reflect the functional connectivity of the network from which the time series data were obtained.

The adjacency matrix is then fed into the Louvain Algorithm, which iterates through different network partitions to maximize network modularity $Q$:

$$Q = \frac{1}{2m} \sum_{ij} \big[A_{ij} - \frac{k_i k_j}{2m}\big] \delta(c_i, c_j) \tag{4}$$

,

where $A_{ij}$ is the adjacency between nodes $i$ and $j$, $k_i$ and $k_j$ are the sums of the adjacencies involving $i$ and $j$, respectively, $c_i$ and $c_j$ are the modules to which nodes $i$ and $j$ have been assigned, respectively, $m = \frac{1}{2} \sum_{ij} A_{ij}$, and $\delta(c_i, c_j)$ equals 1 if $c_i = c_j$ and 0 otherwise.

The Louvain Algorithm for maximizing network modularity alternates between two steps. First, the algorithm optimizes modularity $Q$ locally to find small communities in a network. Then, the algorithm constructs a new network, where each node corresponds to a module detected in the first step. In this new network, connections within modules detected in the first step are represented by a self-loop, and connections across modules detected in the first step are represented by cross-node connections, the weights of which are the sum of the weights of the edges connecting the modules detected in the first step. The Louvain Algorithm then optimizes modularity $Q$ for this new high-level network to find larger-scale communities, and then iterates back to the first step to re-optimize modularity locally. This process is repeated until the algorithm finds the MMP for the entire graph. For details, see [25].

From this point forward $\Phi^{\mathrm{AR}}_{\mathrm{MIB}}$ will refer to $\Phi^{\mathrm{AR}}$ evaluated across the MIB, $\Phi^{\mathrm{AR}}_{\mathrm{MMP}}$ will refer to $\Phi^{\mathrm{AR}}$ evaluated across the MMP, and $\Phi^{\mathrm{AR}}_{\mathrm{atom}}$ will refer to $\Phi^{\mathrm{AR}}$ evaluated accross the atomic partition. Likewise, from this point forward $\Phi^{*}_{\mathrm{MIB}}$ will refer to $\Phi^{*}$ evaluated across the MIB, $\Phi^{*}_{\mathrm{MMP}}$ will refer to $\Phi^{*}$ evaluated across the MMP, and $\Phi^{*}_{\mathrm{atom}}$ will refer to $\Phi^{*}$ evaluated across the atomic partition.

### 2.3 Testing the Performance Of Integrated Information Measures

#### 2.3.1 Simulating Time Series Data Using a Vector Autoregressive Process

To test the performance of the integrated information measures, we first generated multivariate time series data from 15 12-node random networks using a vector autoregressive process with Gaussian noise (see Supplementary Materials S1.2 for details). We repeated this process such that for each of our 15 networks we had 20 ”runs” of time series data.

We wanted to test whether the integrated information measures could detect that a network had been artificially cut (Figure 2). For every intact network that was used to generate time series data, we also introduced every possible cut that splits the network into two equally sized subnetworks.

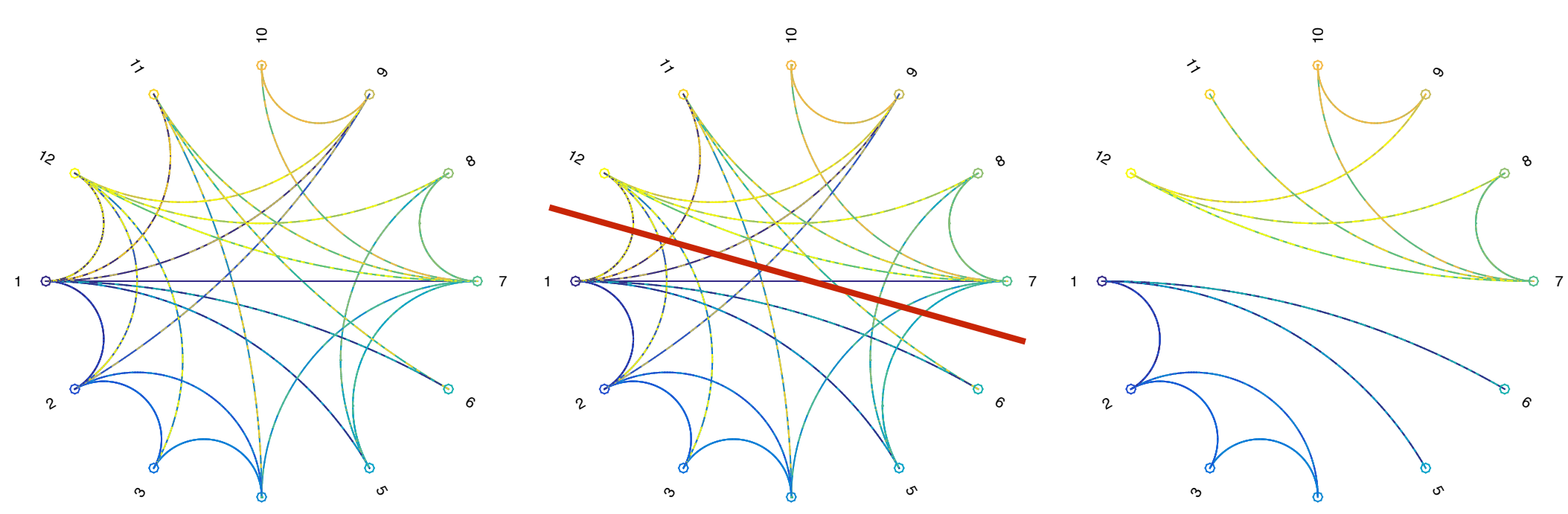

**Figure 2:** Example of one cut. All connections crossing the cut are severed, resulting in two distinct networks.

Using the same vector autoregressive process that was used to generate time series data from the intact networks, we generated 20 runs of time series data for every possible cut for all 15 networks. Because there are 462 possible ways of splitting a 12-node network exactly in half ($\binom{12}{6}/2 = 462$), this gave us 9,240 ”cut” time series for each network (i.e. 20 runs for each of 462 cuts).

### 2.4 Cut Detection Task

If a given measure of integrated information can detect that information is integrated in a network, then it should also detect a lack thereof in a cut network. To test that prediction, we computed all seven integrated information measures – CD, $\Phi^{*}_{\mathrm{MIB}}$, $\Phi^{*}_{\mathrm{MMP}}$, $\Phi^{*}_{\mathrm{atom}}$, $\Phi^{\mathrm{AR}}_{\mathrm{MIB}}$, $\Phi^{\mathrm{AR}}_{\mathrm{MMP}}$, and $\Phi^{\mathrm{AR}}_{\mathrm{atom}}$ – for every run of every network, both intact and cut. $\Phi^{*}$ and $\Phi^{\mathrm{AR}}$, across all partitions, were calculated with a time-lag $\tau$ of 10 time-steps (since these two measures quantify how informative a system is about itself over time - see Supplementary Materials S1.1).

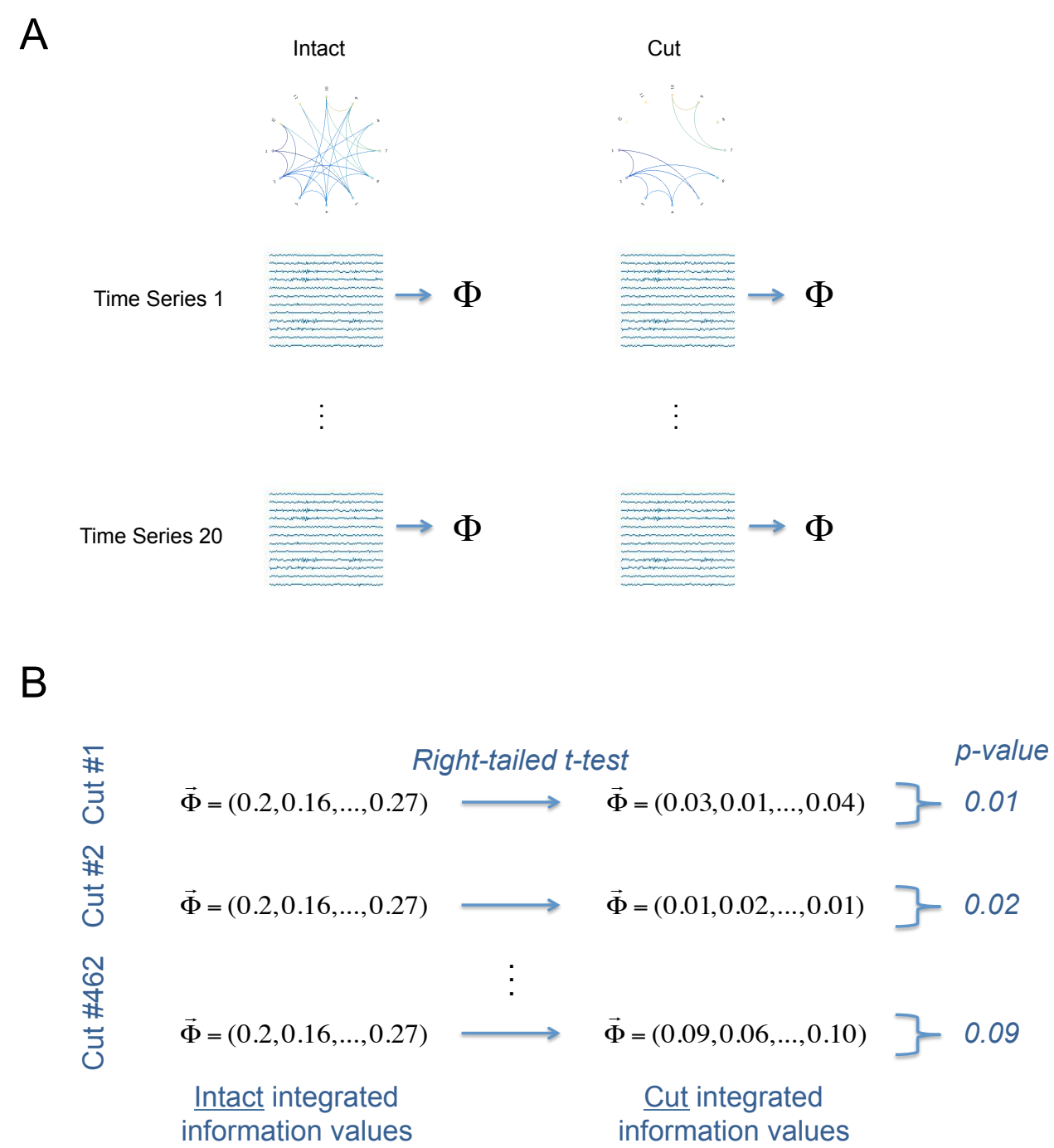

**Figure 3:** Schematic of our detection task. A) For every intact network, we generated 20 runs of multivariate Gaussian time series data. We then severed the network in half, and again generated 20 runs of multivariate Gaussian time series data. Each integrated information measure was calculated for each multivariate time series. B) To assess whether the integrated information measures reliably gave higher values for the intact network than for the cut network, we performed right-tailed t-tests to compare the integrated information values from the intact and cut networks. Because there are 462 ways to split a 12-node network in half ($\binom{12}{6}/2 = 462$), we did this 462 times, once for each cut. The "accuracy" of a measure on a given network was calculated as the percentage of pFDR-corrected p-values that were less than 0.05.

For a given network and a given measure, the above calculations gave us 20 values when the network was intact and 20 estimates of integrated information for each of 462 cuts of that network. To assess whether integrated information was reliably higher in the network when it was intact, we performed a right-tailed two-sample t-test to see if the 20 integrated information values from an intact network were significantly higher than the 20 integrated information values from a given cut of that network (Figure 3). For a single network, this was done for all 462 cuts, resulting in 462 p-values. We then corrected for multiple comparisons (since for a given network, data from each cut were compared to the same intact network) using the positive false discovery rate (pFDR) method [38]. The "accuracy" of a given integrated information measure for that network was then calculated as the percentage of pFDR-corrected p-values that were less than 0.05.

We performed the above "accuracy" test for every integrated information measure for each of our 15 networks, yielding 15 accuracy scores for each measure.

# 3 Results

## 3.1 Detection Performance

Our detection task yielded one accuracy value for each measure, for each network. Each measure's mean accuracy is plotted in Figure 5.

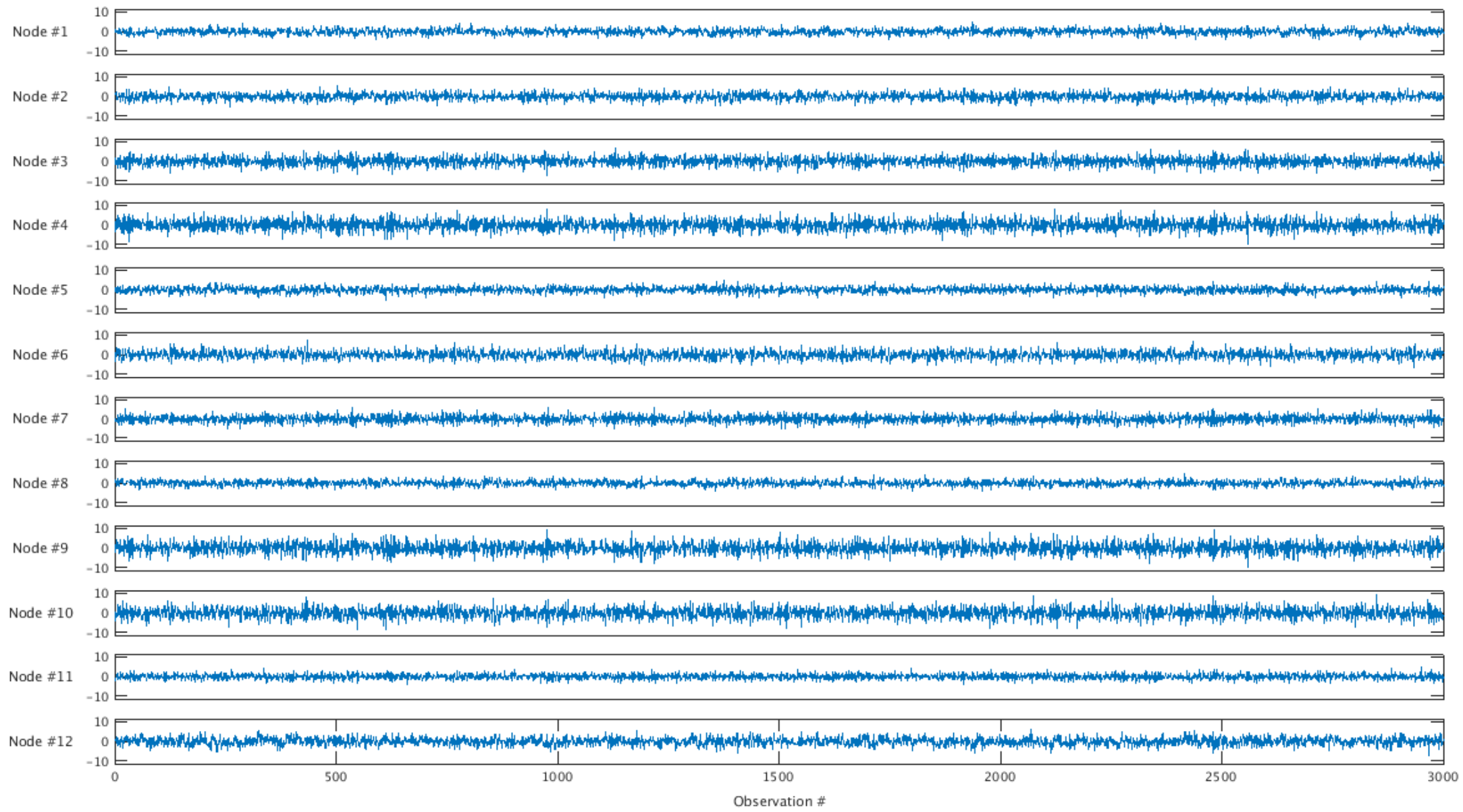

**Figure 4:** A sample multivariate time series from one run of an intact network

Because variances in detection task accuracy varied significantly across measures (Bartlett's test [39], $K^2$=39.207, df=6, p<0.001), detection task performance was compared using the non-parametric Kruskal-Wallis one-way analysis of variance [40]. The Kruskal-Wallis test showed that the mean accuracies of the four measures differed significantly ($\chi^2$=39.263, df=6, p<0.001). Post-hoc pairwise comparisons between the measures' accuracies using right-tailed Wilcoxon signed rank tests showed CD to be significantly more accurate than every other measure (when compared to $\Phi^{\mathrm{AR}}_{\mathrm{MIB}}$: V=120, p<0.001; when compared to $\Phi^{\mathrm{AR}}_{\mathrm{MMP}}$: V=88, p=0.002; when compared to $\Phi^{\mathrm{AR}}_{\mathrm{atom}}$: V=90, p=0.001; when compared to $\Phi^{*}_{\mathrm{MIB}}$: V=105, p<0.001; when compared to $\Phi^{*}_{\mathrm{MMP}}$: V=120, p<0.001; and when compared to $\Phi^{*}_{\mathrm{atom}}$: V=103, p<0.001). Right-tailed Wilcoxon signed rank tests showed $\Phi^{\mathrm{AR}}_{\mathrm{MMP}}$ to be significantly more accurate than $\Phi^{\mathrm{AR}}_{\mathrm{MIB}}$ (V=93, p=0.006), $\Phi^{\mathrm{AR}}_{\mathrm{atom}}$ to be significantly more accurate than $\Phi^{\mathrm{AR}}_{\mathrm{MIB}}$ (V=100, p=0.002), $\Phi^{*}_{\mathrm{MIB}}$ to be significantly more accurate than $\Phi^{*}_{\mathrm{MMP}}$ (V=112, p<0.001), and $\Phi^{*}_{\mathrm{atom}}$ to be significantly more accurate than $\Phi^{*}_{\mathrm{MMP}}$ (V=110, p=0.001). There were no other significant differences in detection task accuracy.

We next looked for variables that could predict detection test accuracy. Of the variables tested, baseline integrated-ness (i.e., how integrated a measure said a network was when it was intact) was the only reliable predictor of a measure's performance on the detection task (see Table S1 for details). The relationship between baseline integrated-ness and detection task accuracy can be seen clearly in the slope of the data plotted in Figure 6.

To further test the relationship between "baseline" integrated information (i.e. integrated information in an intact network) and performance on our detection task, we generated time series data from 100 new 12-node random networks, both intact and cut, following the same procedure as in the previous analysis (see Supplementary Materials S1.2). For these 100 new networks, we only tested $\Phi^{\mathrm{AR}}$ and $\Phi^{*}$ across the MMP and atomic partitions (and not the MIB) to save on computation time (Figure 7). The relationship between baseline integrated-ness and detection task accuracy held, and, for each measure, seems to asymptote at 100% accuracy for more integrated networks (Figure 7).

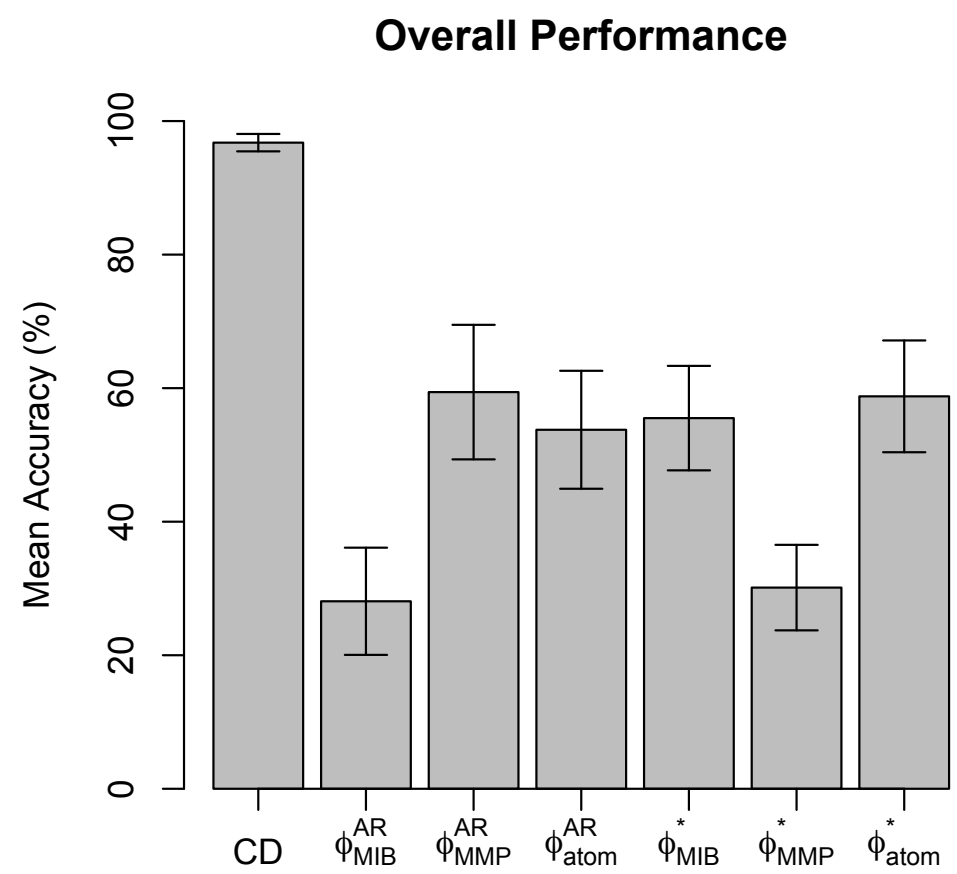

**Figure 5:** Each measure's mean accuracy across 15 networks. Error bars indicate s.e.m.

### 3.2 Computational Speed

Reliability is not the only criterion for choosing which integrated information measure to apply to actual neural data. Leaving aside the *a priori* assumptions of the measures, another important consideration is the time it takes to compute these measures on reasonably sized neural networks. The average computation time for each integrated information measure in the previous analysis is plotted in Figure 8.

To test how network size affects the computation time for these measures, we simulated time series data from networks of varying sizes, following the same vector autoregressive process as in the previous analyses. For each network size, we simulated five networks to get a measure of variance in computation time for a given network size. While the computation time for CD seems to increase exponentially with larger networks (Figure 9), $\Phi^*_{\mathrm{MMP}}$, $\Phi^*_{\mathrm{atom}}$, $\Phi^{\mathrm{AR}}_{\mathrm{MMP}}$, and $\Phi^{\mathrm{AR}}_{\mathrm{atom}}$ are computed in less than five seconds for even the largest networks tested (Figure 9). Because of the inhomogeneity of their variances (Bartlett's test, $K^2$=731.1, df=4,p<0.001), the average run times were compared using a Kruskal-Wallis rank sum test. The test showed a significant difference in the effect of network size on run time for the five measures ($\chi^2$=94.082,df=4,p<0.001). Post-hoc pairwise comparisons between run times using right-tailed Wilcoxon signed rank tests showed that CD is significantly slower to compute than $\Phi^*_{\mathrm{MMP}}$ (V=275, p<0.001), $\Phi^*_{\mathrm{atom}}$ (V=271, p<0.001), $\Phi^{\mathrm{AR}}_{\mathrm{MMP}}$ (V=276, p<0.001), and $\Phi^{\mathrm{AR}}_{\mathrm{atom}}$ (V=276, p<0.001), that $\Phi^*_{\mathrm{atom}}$ is significantly slower to compute than $\Phi^*_{\mathrm{MMP}}$ (V=275,p<0.001), that $\Phi^*_{\mathrm{atom}}$ is significantly slower to compute than $\Phi^{\mathrm{AR}}_{\mathrm{MMP}}$ (V=276,p<0.001), that $\Phi^*_{\mathrm{atom}}$ is significantly slower to compute than $\Phi^{\mathrm{AR}}_{\mathrm{atom}}$ (V=276,p<0.001), that $\Phi^*_{\mathrm{MMP}}$ is significantly slower to compute than $\Phi^{\mathrm{AR}}_{\mathrm{MMP}}$ (V=276,p<0.001), that $\Phi^*_{\mathrm{MMP}}$ is significantly slower to compute than $\Phi^{\mathrm{AR}}_{\mathrm{atom}}$ (V=276,p<0.001), and that $\Phi^{\mathrm{AR}}_{\mathrm{MMP}}$ is significantly slower to compute than $\Phi^{\mathrm{AR}}_{\mathrm{atom}}$ (V=276,p<0.001).

### 3.3 Correlations Between Measures

Finally, we wanted to see how well integrated information measures across different network partitions track each other. If different measures produce similar values across different networks, then a measure's reliability, computation speed, and underlying *a priori* assumptions should be the primary decision criteria when considering which measure to use on real neural data. On the other hand, if the measures do not produce similar results across different networks, then it becomes less clear how to pick a measure to use on real neural data.

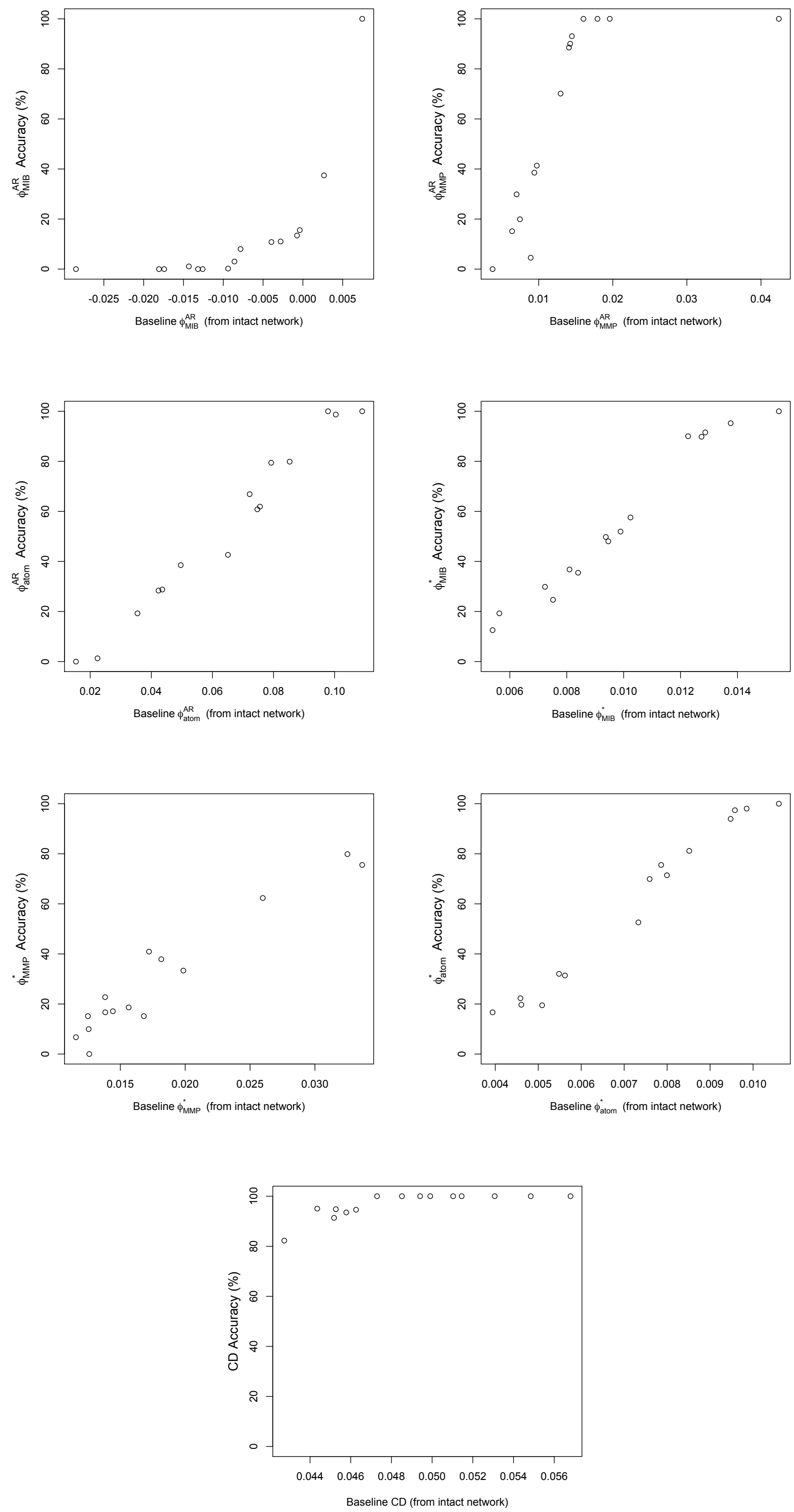

**Figure 6:** Although Causal Density is on average the most accurate measure, the other integrated information measures reliably detect cuts in networks that are more integrated to begin with. In each figure, the dots represent individual random 12-node networks.

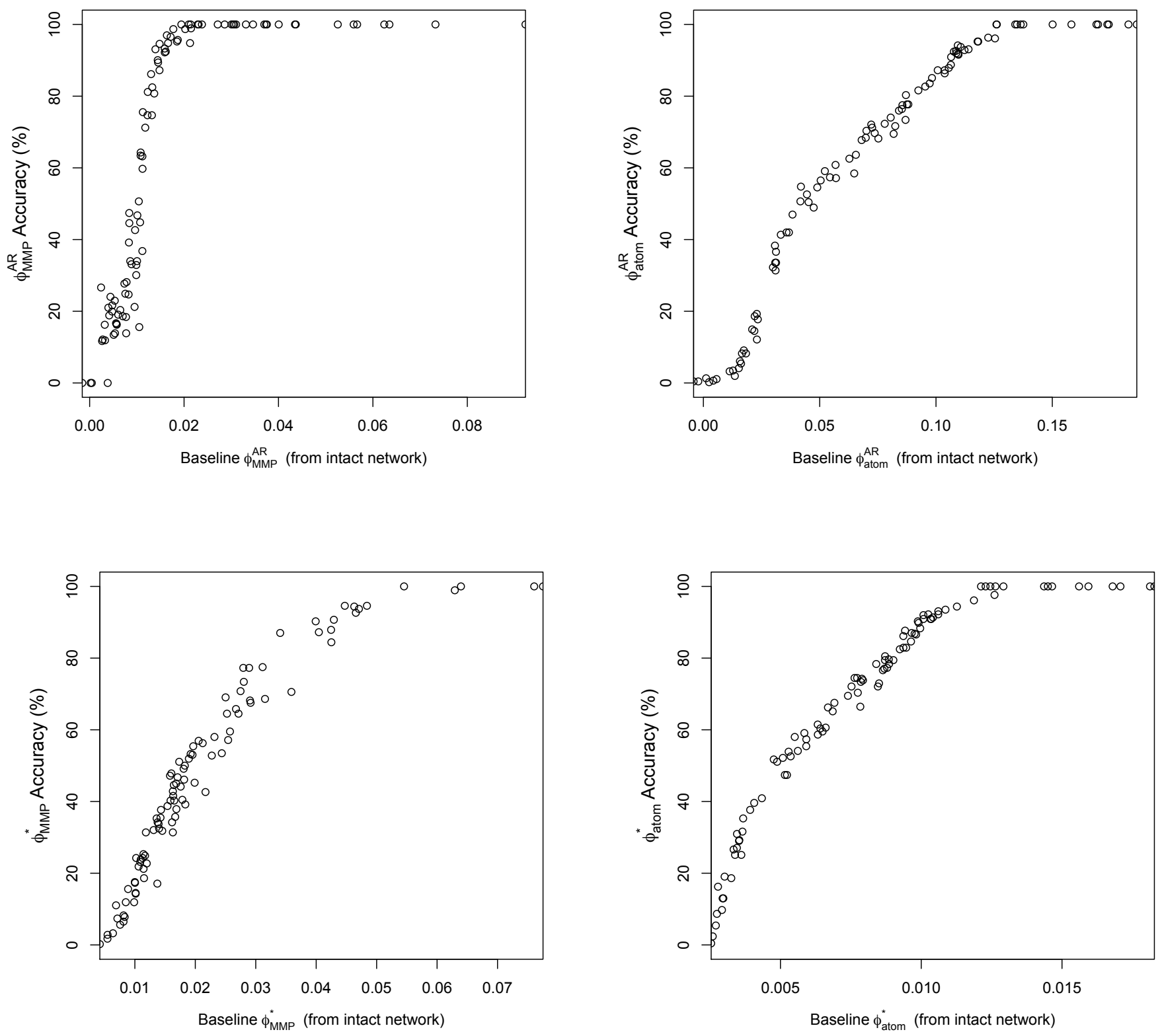

**Figure 7:** The relationship between baseline integrated-ness and detection accuracy for $\Phi^{\mathrm{MMP}}$ and $\Phi^{*}$, across both the MMP and atomic partition, asymptote around 100% accuracy. Each dot represents one out of 100 new networks simulated to test the relationship between baseline integrated-ness and detection task accuracy.

To compare different measures of integrated information to each other, we computed the mean of each measure across 20 runs of time series data generated from all of our intact networks. For the 15 networks plotted in Figures 5 and 6, we were able to compute the mean value of all seven integrated information measures. For the 100 networks plotted in Figure 7, we only computed the five fast integrated information measures: CD, $\Phi^{\mathrm{AR}}_{\mathrm{MMP}}$, $\Phi^{\mathrm{AR}}_{\mathrm{atom}}$, $\Phi^{*}_{\mathrm{MMP}}$, and $\Phi^{*}_{\mathrm{atom}}$. We then took the Pearson's correlation *r* between all seven measures in the 15 networks plotted in Figures 5 and 6 (Table 1) and between the five fast measures in the 100 networks plotted in Figure 7 (Table 2). We found that although the correlations were weaker when we only considered 15 networks, strong correlations emerged between all measures across 100 networks. We found that across the 15 networks for which we calculated every integrated information measure, the only correlations that were significant (after FDR correction) were between $\Phi^{*}_{\mathrm{MMP}}$ and $\Phi^{\mathrm{AR}}_{\mathrm{MMP}}$ (p=0.005), $\Phi^{*}_{\mathrm{MIB}}$ and $\Phi^{\mathrm{AR}}_{\mathrm{atom}}$ (p<0.001), $\Phi^{*}_{\mathrm{atom}}$ and $\Phi^{\mathrm{AR}}_{\mathrm{atom}}$ (p<0.001), and $\Phi^{*}_{\mathrm{atom}}$ and $\Phi^{*}_{\mathrm{MIB}}$ (p<0.001) (Table 1). Across 100 networks, however, all five of the fast measures of integrated information were significantly correlated with one another (Table 2).

## 4 Discussion

In this study, we introduced a partition that has been studied extensively in network neuroscience, the Maximum Modularity Partition (MMP), to Integrated Information Theory. In combination with previously proposed measures, $\Phi^{*}$ and $\Phi^{\mathrm{AR}}$, the MMP yields two new measures of integrated information: $\Phi^{*}_{\mathrm{MMP}}$ and $\Phi^{\mathrm{AR}}_{\mathrm{MMP}}$. We compared $\Phi^{*}_{\mathrm{MMP}}$ and $\Phi^{\mathrm{AR}}_{\mathrm{MMP}}$ to $\Phi^{*}$ and $\Phi^{\mathrm{AR}}$ across previously proposed network partitions, and also compared the two measures to Causal Density (CD).

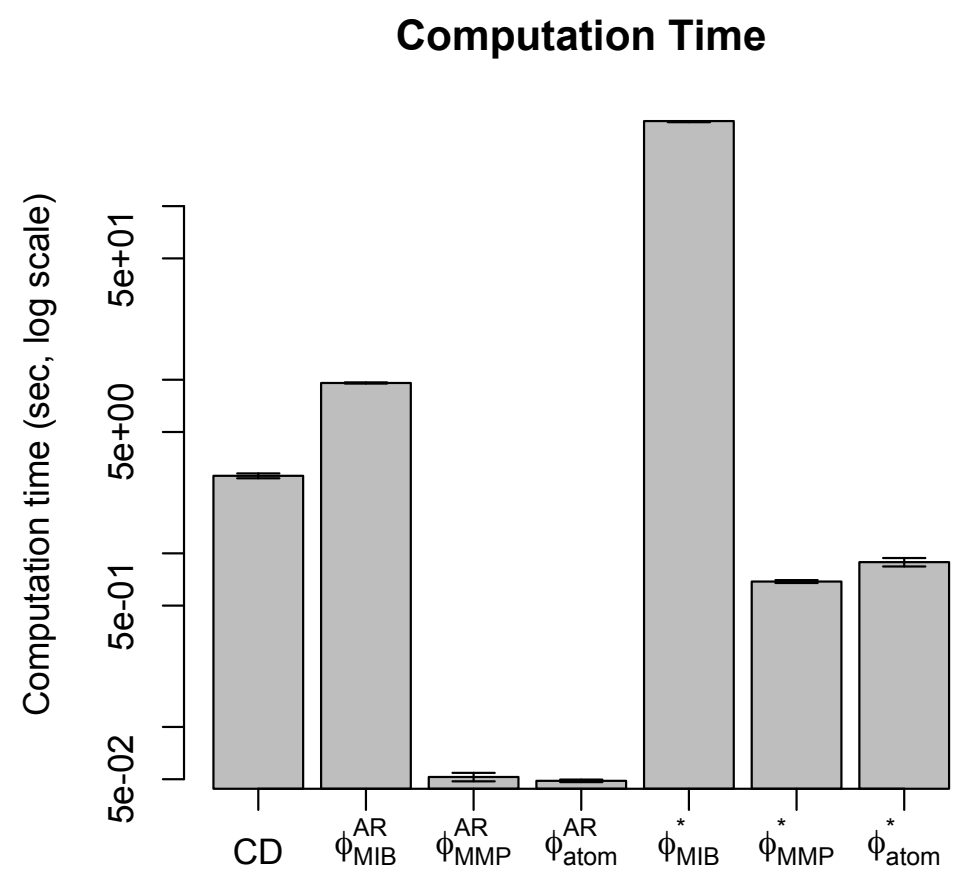

**Figure 8:** Each measure's mean computation time across 15 networks, plotted on a logarithmic scale. For each network, computation time was measured as the average run time for 20 runs of the network when it was intact. Error bars indicate s.e.m. across the 15 networks.

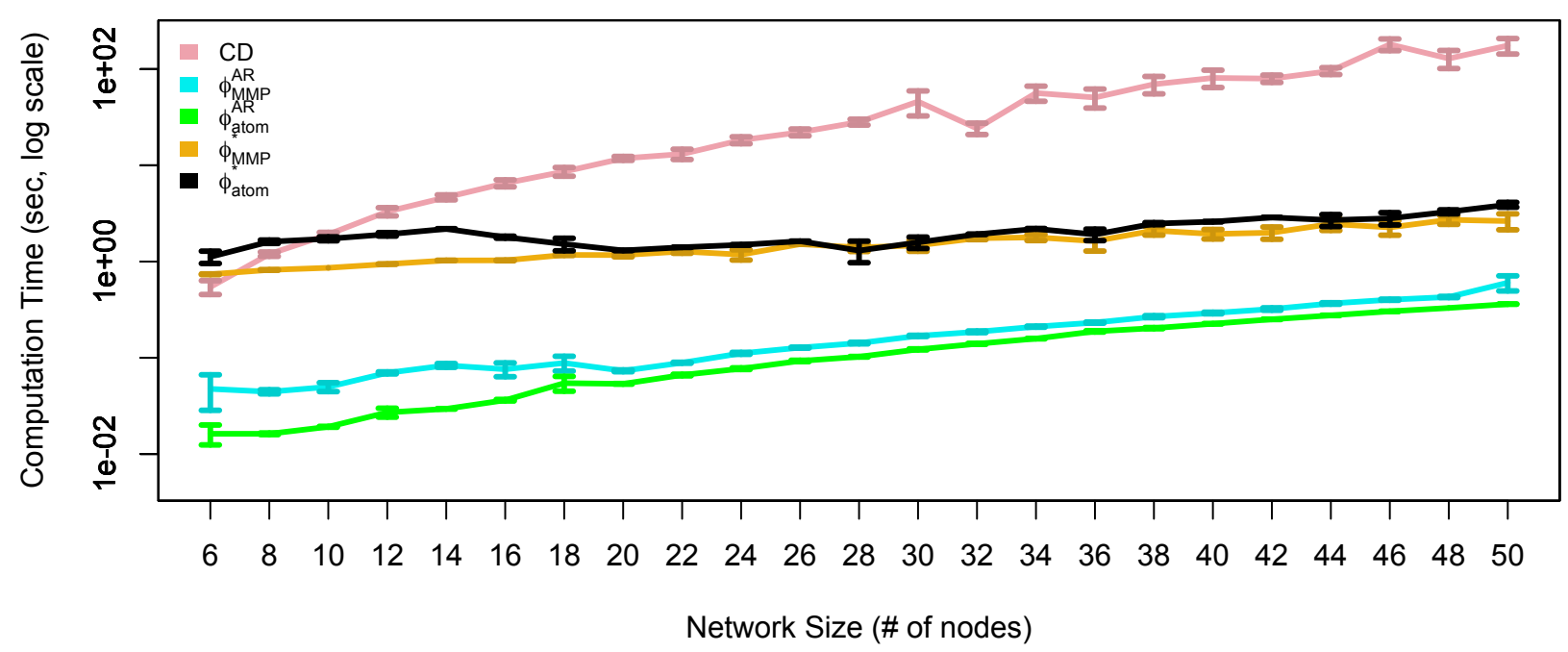

**Figure 9:** Network size vs. run time for CD, $\Phi^{\rm AR}_{\rm MMP}$, $\Phi^{\rm AR}_{\rm atom}$, $\Phi^{*}_{\rm MMP}$, and $\Phi^{*}_{\rm atom}$, plotted on a logarithmic scale. The number of edges in each network was scaled proportionately to the size of the network, such that there were four times the number of edges as there were nodes. Error bars indicate s.e.m. across the five networks simulated for a given network size. Note that computation time for CD rapidly increases in relation to network size, whereas $\Phi^{\rm AR}$ and $\Phi^{*}$ across both the MMP and the atomic partition remain fast even at larger network sizes.

We found that in a detection task that tested whether a measure could detect a structural bipartition of a network, CD outperformed the measures of integrated information. While CD was on average more accurate than the other measures tested, the measures of integrated information could reliably detect cuts in more integrated networks than they could in less integrated networks (Figures 6 and 7). Although this pattern held for both $\Phi^{*}$ and $\Phi^{\rm AR}$ across all three partitions tested - the MIB, the MMP, and the atomic partition - the choice of partition did significantly affect the two measures' average accuracy: $\Phi^{\rm AR}$ performed significantly better on average across the MMP and the atomic partition than it did across the MIB, and $\Phi^{*}$ performed significantly better across the MIB and the atomic partition than across the MMP (Figure 5).

It is worth pointing out that our detection task gave a natural advantage to the MIB over the MMP and the atomic partition, because it assessed how well the measures could detect a structural *bi*partition of a network. As such, it is surprising that, on average, $\Phi^{\rm AR}$ could more reliably detect a bipartition when evaluated across the MMP or atomic partition than across the MIB, and that $\Phi^*$ could detect a bipartition as reliably when calculated across the atomic partition as it could when evaluated across the MIB.

| | CD | $\phi^{AR}_{MIB}$ | $\phi^{AR}_{MMP}$ | $\phi^{AR}_{atom}$ | $\phi^{*}_{MIB}$ | $\phi^{*}_{MMP}$ | $\phi^{*}_{atom}$ |
|---|---|---|---|---|---|---|---|
| CD | 1 | | | | | | |
| $\phi^{AR}_{MIB}$ | 0.28 | 1 | | | | | |
| $\phi^{AR}_{MMP}$ | 0.22 | 0.59 | 1 | | | | |
| $\phi^{AR}_{atom}$ | 0.12 | 0.1 | 0.44 | 1 | | | |
| $\phi^{*}_{MIB}$ | 0.26 | 0.03 | 0.51 | **0.84*** | 1 | | |
| $\phi^{*}_{MMP}$ | 0.06 | 0.26 | **0.77** | 0.44 | 0.58 | 1 | |
| $\phi^{*}_{atom}$ | 0.13 | -0.04 | 0.39 | **0.96*** | **0.91*** | 0.44 | 1 |

**Table 1:** Pearson's correlation *r* between all measures of integrated information evaluated across the 15 networks analyzed in Figures 5 and 6. Bolded correlation values indicate significance at an alpha of 0.05 after FDR correction ($*$: 0.01>p<0.05, $**$: 0.001>p<0.01, ***: p<0.001).

| | CD | $\phi^{AR}_{MMP}$ | $\phi^{AR}_{atom}$ | $\phi^{*}_{MMP}$ | $\phi^{*}_{atom}$ |
|---|---|---|---|---|---|
| CD | 1 | | | | |
| $\phi^{AR}_{MMP}$ | **0.32** | 1 | | | |
| $\phi^{AR}_{atom}$ | **0.53*** | **0.65*** | 1 | | |
| $\phi^{*}_{MMP}$ | **0.24*** | **0.88*** | **0.60*** | 1 | |
| $\phi^{*}_{atom}$ | **0.54*** | **0.65*** | **0.96*** | **0.67*** | 1 |

**Table 2:** Pearson's correlation *r* between fast measures of integrated information evaluated across the 100 networks analyzed in Figure 7. Bolded correlation values indicate significance at an alpha of 0.05 after FDR correction ($*$: 0.01>p<0.05, $**$: 0.001>p<0.01, ***: p<0.001).

We also found that when evaluated on larger networks, $\Phi^*$ and $\Phi^{\mathrm{AR}}$ across either the MMP or atomic partition computed in under five seconds, while the computation time for CD rapidly increased in proportion to network size (Figure 9).

Finally, we found that although only a few of the measures correlated with each other significantly across the 15 networks plotted in Figure 6 (Table 1), significant correlations between all five of the faster measures - CD, $\Phi^{\mathrm{AR}}_{\mathrm{MMP}}$, $\Phi^{\mathrm{AR}}_{\mathrm{atom}}$, $\Phi^{*}_{\mathrm{MMP}}$, and $\Phi^{*}_{\mathrm{atom}}$ - emerged across the 100 networks plotted in Figure 7 (Table 2).

Taken together, these results have several implications for how to calculate integrated information from time series data in practice. Although CD was the most accurate measure and also correlated with the other measures, it has several drawbacks: CD can only be applied to Gaussian data, it is not strictly speaking a measure of integrated information, and its computation time increases rapidly with larger networks. That said, if Gaussianity can be assumed, then CD is a worthwhile measure to consider.

If a researcher wants to use a measure of integrated information in the stricter sense (i.e., assess information in a system that is greater than in the sum of the parts), then $\Phi^*$ is the best measure to apply if Gaussianity can be assumed, since it is more theoretically sound than $\Phi^{\mathrm{AR}}$. Specifically, as in [10], $\Phi^{\mathrm{AR}}$ produced negative values in this study (Figures 6 and 7). This is not a surprise, because $\Phi^{\mathrm{AR}}$ is based on differential entropy, which is known to suffer from the possibility of being negative. As [10] and [11] point out, integrated information cannot be negative, and as such $\Phi^{\mathrm{AR}}$ violates a key assumption. Because we showed that $\Phi^*$ is highly correlated with $\Phi^{\mathrm{AR}}$ (Tables 1 and 2), performs as well as $\Phi^{\mathrm{AR}}$ on our detection task (Figures 5, 6, and 7), and is essentially as fast to compute as $\Phi^{\mathrm{AR}}$ across the MMP and atomic partition (Figures 8, 9, and 10), we recommend calculating $\Phi^*$ across the MMP or atomic partition when Gaussianity can be assumed or when data can be transformed into a normal distribution.

That said, normality cannot always be assumed, nor can neural data always be transformed into a normal distribution. This is the case, for example, with spike-timing data [31]. For such data, $\Phi^{\mathrm{AR}}$ may be a useful measure of integrated information because, unlike $\Phi^*$, it does not require calculating empirical covariance matrices, which, for accurate estimation in non-Gaussian variables, may require more data than is practical to obtain.

To our knowledge, the MMP has not been considered in the context of Integrated Information Theory before. We showed that calculating integrated information across the MMP or the atomic partition, which was used in [10], has the clear advantage over the MIB in being generally as accurate and much faster to compute. Although we do not have empirical evidence to assess the merits of the MMP over the atomic partition, it is worth reiterating that the MMP is more in line with what is known about large-scale information flow in the brain: the brain follows a broadly modular architecture in both its anatomy [14], [15] and functional connectivity [15], [19], [20], [21], and information in the brain seems to be integrated across relatively discrete functional modules [22], [21]. For these reasons, evaluating integrated information across the MMP is more congruent with what previous work has shown regarding large-scale information flow in the brain than is evaluating integrated information across the atomic partition.

Because calculating integrated information across the MMP or the atomic partition is orders of magnitude faster than calculating integrated information across the MIB - without loss of accuracy for more integrated networks - evaluating integrated information across either of these two partitions represents a significant step forward in the aim of calculating integrated information in real brains.

**Acknowledgments**

We would like to thank the members of Mark D'Esposito's laboratory at UC Berkeley for help with our graph-theoretic analyses, and we would also like to thank the members of UC Berkeley's Redwood Center for Theoretical Neuroscience for their helpful feedback.

## References

[1] Tononi G. An information integration theory of consciousness. BMC neuroscience. 2004;5(1):42.

[2] Tononi G. Consciousness as integrated information: a provisional manifesto. The Biological Bulletin. 2008;215(3):216–242.

[3] Tononi G. Integrated information theory of consciousness: an updated account. Arch Ital Biol. 2012;150(2-3):56–90.

[4] Oizumi M, Albantakis L, Tononi G. From the phenomenology to the mechanisms of consciousness: integrated information theory 3.0. PLoS Comput Biol. 2014;10(5):e1003588.

[5] Searle JR. Can information theory explain consciousness. New York Review of Books. 2013;p. 54–58.

[6] Walker SI. Top-down causation and the rise of information in the emergence of life. Information. 2014;5(3):424–439.

[7] Seth AK, Izhikevich E, Reeke GN, Edelman GM. Theories and measures of consciousness: an extended framework. Proceedings of the National Academy of Sciences. 2006;103(28):10799–10804.

[8] Seth AK, Barrett AB, Barnett L. Causal density and integrated information as measures of conscious level. Philosophical Transactions of the Royal Society of London A: Mathematical, Physical and Engineering Sciences. 2011;369(1952):3748–3767.

[9] Barrett AB, Seth AK. Practical measures of integrated information for time-series data. PLoS Comput Biol. 2011;7(1):e1001052.

[10] Oizumi M, Amari Si, Yanagawa T, Fujii N, Tsuchiya N. Measuring integrated information from the decoding perspective. PLoS Comput Biol. 2016;12(1):e1004654.

[11] Tegmark M. Improved Measures of Integrated Information. arXiv preprint arXiv:160102626. 2016;.

[12] Ihmels J, Friedlander G, Bergmann S, Sarig O, Ziv Y, Barkai N. Revealing modular organization in the yeast transcriptional network. Nature genetics. 2002;31(4):370–377.

[13] Guimera R, Amaral LAN. Functional cartography of complex metabolic networks. Nature. 2005;433(7028):895–900.

[14] Chen ZJ, He Y, Rosa-Neto P, Germann J, Evans AC. Revealing modular architecture of human brain structural networks by using cortical thickness from MRI. Cerebral cortex. 2008;18(10):2374–2381.

[15] Bullmore E, Sporns O. Complex brain networks: graph theoretical analysis of structural and functional systems. Nature Reviews Neuroscience. 2009;10(3):186–198.

[16] Oldham MC, Konopka G, Iwamoto K, Langfelder P, Kato T, Horvath S, et al. Functional organization of the transcriptome in human brain. Nature neuroscience. 2008;11(11):1271–1282.

[17] Hawrylycz MJ, Lein ES, Guillozet-Bongaarts AL, Shen EH, Ng L, Miller JA, et al. An anatomically comprehensive atlas of the adult human brain transcriptome. Nature. 2012;489(7416):391–399.

[18] Richiardi J, Altmann A, Milazzo AC, Chang C, Chakravarty MM, Banaschewski T, et al. Correlated gene expression supports synchronous activity in brain networks. Science. 2015;348(6240):1241–1244.

[19] Bruno AM, Frost WN, Humphries MD. Modular deconstruction reveals the dynamical and physical building blocks of a locomotion motor program. Neuron. 2015;86(1):304–318.

[20] Power JD, Cohen AL, Nelson SM, Wig GS, Barnes KA, Church JA, et al. Functional network organization of the human brain. Neuron. 2011;72(4):665–678.

[21] Bertolero MA, Yeo BT, D'Esposito M. The modular and integrative functional architecture of the human brain. Proceedings of the National Academy of Sciences. 2015;112(49):E6798–E6807.

[22] van den Heuvel MP, Sporns O. Network hubs in the human brain. Trends in cognitive sciences. 2013;17(12):683–696.

[23] Girvan M, Newman ME. Community structure in social and biological networks. Proceedings of the national academy of sciences. 2002;99(12):7821–7826.

[24] Newman ME, Girvan M. Finding and evaluating community structure in networks. Physical review E. 2004;69(2):026113.

[25] Blondel VD, Guillaume JL, Lambiotte R, Lefebvre E. Fast unfolding of communities in large networks. Journal of statistical mechanics: theory and experiment. 2008;2008(10):P10008.

[26] Ovelgönne M, Geyer-Schulz A. An ensemble learning strategy for graph clustering. Graph Partitioning and Graph Clustering. 2012;588:187.

[27] Alvarez AJ, Sanz-Rodríguez CE, Cabrera JL. Weighting dissimilarities to detect communities in networks. Phil Trans R Soc A. 2015;373(2056):20150108.

[28] Balduzzi D, Tononi G. Integrated information in discrete dynamical systems: motivation and theoretical framework. PLoS Comput Biol. 2008;4(6):e1000091.

[29] Tononi G, Sporns O. Measuring information integration. BMC neuroscience. 2003;4(1):31.

[30] Tononi G. Consciousness, information integration, and the brain. Progress in brain research. 2005;150:109–126.

[31] Rieke F, Warland D, de Ruyter van Steveninck R, Bialek W. MIT Press; Cambridge, MA: 1996. Spikes: exploring the neural code, a Bradford book;.

[32] Van Albada S, Robinson P. Transformation of arbitrary distributions to the normal distribution with application to EEG test–retest reliability. Journal of neuroscience methods. 2007;161(2):205–211.

[33] Baker J, Weisskoff R, Stern C, Kennedy D, Jiang A, Kwong K, et al. Statistical assessment of functional MRI signal change. In: Proceedings of the 2nd Annual Meeting of the Society of Magnetic Resonance. vol. 626; 1994. .

[34] Wu D, Lewin J. Evaluation of non-parametric statistical measures and data clustering for functional MR data analysis. In: Proceedings of the SMR 2nd Annual Meeting; 1994. p. 629.

[35] Merhav N, Kaplan G, Lapidoth A, Shitz SS. On information rates for mismatched decoders. Information Theory, IEEE Transactions on. 1994;40(6):1953–1967.

[36] Granger CW. Investigating causal relations by econometric models and cross-spectral methods. Econometrica: Journal of the Econometric Society. 1969;p. 424–438.

[37] Sporns O. Networks of the Brain. MIT press; 2011.

[38] Storey JD, Tibshirani R. Statistical significance for genomewide studies. Proceedings of the National Academy of Sciences. 2003;100(16):9440–9445.

[39] Bartlett MS. Properties of sufficiency and statistical tests. Proceedings of the Royal Society of London Series A, Mathematical and Physical Sciences. 1937;p. 268–282.

[40] Kruskal WH, Wallis WA. Use of ranks in one-criterion variance analysis. Journal of the American statistical Association. 1952;47(260):583–621.

[41] Seth AK. A MATLAB toolbox for Granger causal connectivity analysis. Journal of neuroscience methods. 2010;186(2):262–273.

[42] Barnett L, Barrett AB, Seth AK. Granger causality and transfer entropy are equivalent for Gaussian variables. Physical review letters. 2009;103(23):238701.

[43] Erdős P, Rényi A. On random graphs. Publicationes Mathematicae Debrecen. 1959;6:290–297.

[44] Barnett L, Seth AK. The MVGC multivariate Granger causality toolbox: a new approach to Granger-causal inference. Journal of neuroscience methods. 2014;223:50–68.

[45] Royston J. Some techniques for assessing multivarate normality based on the shapiro-wilk w. Applied Statistics. 1983;p. 121–133.

[46] Bonferroni CE. Teoria statistica delle classi e calcolo delle probabilita. Libreria internazionale Seeber; 1936.

# S1 Supplementary Materials

## S1.1 Measures of Integrated Information (mathematical details)

### $\Phi^{\mathrm{AR}}$

Computing $\Phi^{\mathrm{AR}}$ [9] begins with simple linear regressions. For two random variables $X$ and $Y$, the linear regression of $X$ on $Y$ can be calculated as follows:

$$X = \alpha + A{\cdot}Y + E \tag{S1}$$

where $\alpha$ is a vector of constants, $A$ is the regression matrix, and $E$ is the prediction error. Barrett and Seth utilize the fact that for any variables $X$ and $Y$, even when not Gaussian,

$$\Sigma(E) = \Sigma(X|Y) \tag{S2}$$

In other words, the covariance matrix of the prediction errors from the linear regression is equivalent to the partial covariance of $X$ given $Y$.

With this equivalence in hand, Barrett and Seth introduce a measure of integrated information, based on the intuition that integrated information should be the information in the whole of a system that is greater than the sum of its parts.

To calculated integrated information, Barrett and Seth begin with the following linear regressions:

$$X_{t-\tau} = A^{X}{\cdot}X_t + E_t^{X} \tag{S3}$$

and

$$M^k_{t-\tau} = A^{M^k}{\cdot}M^k_t + E_t^{M^k} \tag{S4}$$

where $X$ is the whole system, $M^k$ is a sub-system of $X$, $t$ is the present, and $t - \tau$ is the past. In Eq. S3, the past of $X$ is regressed on its present, and in Eq. S4, the past of a sub-system $M^k$ of $X$ is regressed on its own present. The prediction errors in these regressions are then used to calculate "effective information" across *n* sub-systems of $X$, relying on the equivalence in Equation S2:

$$\varphi^{\mathrm{AR}}[X;\tau,\{M^1, M^2, ..., M^n\}] =: \frac{1}{2}\mathrm{log}\left\{\frac{\mathrm{det}\Sigma(X)}{\mathrm{det}\Sigma(E^X)}\right\} - \sum_{k=1}^{n}\frac{1}{2}\mathrm{log}\left\{\frac{\mathrm{det}\Sigma(M^k)}{\mathrm{det}\Sigma(E^{M^k})}\right\} \tag{S5}$$

When evaluated across the MIB, the *n* in Equation S5 is 2. When evaluated across the MMP, *n* is the number of modules that the Lovain Algorithm (or another modularity optimization algorithm) picks out as giving the MMP of the network. When evaluated across the atomic partition, *n* is the number of nodes in the network.

$\varphi^{\mathrm{AR}}$ is then used to calculate integrated information. When evaluated across either the MMP or the atomic partition, integrated information for some system $X$, across a partition $P$, and over a given time-lag $\tau$ is calculated as:

$$\Phi^{\mathrm{AR}}[X;\tau;P] =: \frac{\varphi^{\mathrm{AR}}[X;\tau;P]}{L(P)} \tag{S6}$$

where the normalizaiton factor $L$ for some partition $P$ is:

$$L(M^1, M^2, ..., M^n) =: \frac{1}{2}\mathrm{logmin}_k(2\pi e)^{|M^k|}\mathrm{det}\Sigma(M^k) \tag{S7}$$

When evaluated across the MIB, integrated information for some system $X$ over a given time-lag $\tau$ is calculated as:

$$\Phi^{\mathrm{AR}}_{\mathrm{MIB}}[X;\tau] =: \arg_P \min \frac{\varphi^{\mathrm{AR}}[X;\tau;P]}{L(P)} \quad \text{(S8)}$$

In other words, to calculate $\Phi^{\mathrm{AR}}_{\mathrm{MIB}}$, evaluate $\varphi^{\mathrm{AR}}$ normalized by $L(P)$ across every bipartition. The lowest value that this produces is $\Phi^{\mathrm{AR}}_{\mathrm{MIB}}$.

### $\Phi^*$

Oizumi et al [10] calculate integrated information as

$$\Phi^*(X^{t-\tau};X^{\tau}) = I(X^{t-\tau};X^{\tau}) - I^*(X^{t-\tau};X^{\tau}) \quad \text{(S9)}$$

where $t$ is the present of the system, $t-\tau$ is the past of the system, $I$ is the mutual information between the past and present of the intact system, and $I^*$ is the mutual information between the past and present of the system when it is partitioned.

Formally, $I$ is calculated as:

$$I(X^{t-\tau};X^{\tau}) = -\sum_{X^t} p(X^t)\log p(X^t) + \sum_{X^{t-\tau},X^{\tau}} p(X^{t-\tau},X^{\tau})\log p(X^t|X^{t-\tau}) \quad \text{(S10)}$$

and $I*$ is calculated as:

$$I^*_{\mathrm{partition}}(X^{t-\tau};X^{\tau}) = -\sum_{X^t} p(X^t)\log \sum_{X^{t-\tau}} p(X^{t-\tau})q(X^t|X^{t-\tau})^{\beta} + \sum_{X^{t-\tau},X^{\tau}} p(X^{t-\tau},X^{\tau})\log q(X^t|X^{t-\tau})^{\beta}, \quad \text{(S11)}$$

$\beta$ is the value that maximizes $I^*(X^{t-\tau};X^{\tau})$, and can be calculated by differentiating $I^*(X^{t-\tau};X^{\tau})$ and solving the equation $dI^*(\beta)/d\beta = 0$ using gradient descent. $p(X^t|X^{t-\tau})$ is the "true" conditional distribution between the past and present states of a system, and $q(X^t|X^{t-\tau})$ is the "partitioned" or "mismatched" probability distribution between the past and present state of the system. The mismatched probability distribution will depend on the partition used. (For more details, see [10]).

### Causal Density

Seth et al (2006) define Causal Density (CD) as the average pairwise Granger causality between every node in a network.

A common framework for calculating Granger causality is based on linear autoregression. First, a variable $X$ is regressed on $p$ lags of itself, conditioned on $r$ lags of variable $Z$:

$$X_t = A(X^{(p)}_{t-1} \oplus Z^{(r)}_{t-1}) + \varepsilon_t \quad \text{(S12)}$$

Then, $X$ is regressed on $p$ lags of itself, but conditioned both on $r$ lags of $Z$ and on $q$ lags of the predictor variable $Y$:

$$X_t = A'(X^{(p)}_{t-1} \oplus Y^{(p)}_{t-1} \oplus Z^{(r)}_{t-1}) + \varepsilon'_t \quad \text{(S13)}$$

For the above regressions, $p$, $q$, and $r$ can be selected using the Akaike or Bayesian information criterion [41].

The Granger causality from $Y$ to $X$ conditioned on $Z$ is then calculated by comparing the residual variances in the two autoregression models:

$$F_{Y\to X|Z} =: \ln\left(\frac{\mathrm{AR}(\varepsilon_t)}{\mathrm{AR}(\varepsilon_t')}\right) = \ln\left(\frac{\Sigma(\varepsilon_t)}{\Sigma(\varepsilon_t')}\right) = \ln\left(\frac{\Sigma(X|X^-\oplus Z^-)}{\Sigma(X|X^-\oplus Y^-\oplus Z^-)}\right) \tag{S14}$$

It is worthwhile to note that for Gaussian variables, there exists a linear relationship between Granger causality and transfer entropy [42], such that

$$F_{Y\to X|Z} = 2F_{T\to X|Z} \tag{S15}$$

which implies that in Gaussian data, there is a direct relationship between Granger causality and information flow.

CD in a system $X$ with $n$ nodes is then calculated as:

$$CD(X) =: \frac{1}{n(n-1)}\sum_{i\neq j} F_{X_i\to X_j|X_{[ij]}} \tag{S16}$$

### S1.2 Generating Multivariate Gaussian Time Series Data

To test whether measures of integrated information could detect cuts in a network, we first created 15 random 12-node networks with 50 edges using the Erdős-Rényi model [43]. To generate time series data from our graphs, we first transformed each binary adjacency matrix into a three-dimensional weighted adjacency matrix with a spectral radius of 0.9. We did this by first creating two matrices with random numbers where there was a 1 in the original adjacency matrix, concatenating the two matrices into a three-dimensional matrix, and then using the var_specrad function of the MVGC Multivariate Granger Causality Toolbox [44] to ”decay” the coefficients in the three-dimensional adjacency matrix such that its spectral radius was 0.9. This process is important for ensuring that activity in the simulated time series does not “explode” into high values.

Using the var_to_tsdata function from the MVGC Multivariate Granger Causality Toolbox [44], the decayed three-dimensional adjacency matrix was then used to generate a multivariate time series with 3000 time-points and a model order of 2 (the model order of the time series is fixed by the fact that the size of the adjacency matrix along the third dimension is 2).

### S1.3 Confirming that Time Series Data Are Multivariate Gaussian

A sample multivariate time series is shown in Figure 4. To test whether the time series data in our analyses were in fact multivariate Gaussian - which is a necessary assumption for computing both CD and $\Phi^*$ - we applied Royston’s Multivariate Normality Test [45] to each time series. Before correcting for multiple comparisons, Royston’s test failed to reject the null hypothesis of non-normality at an alpha of 0.05 for every time series from the intact networks (suggesting that they were in fact Gaussian), but did reject the null hypothesis of non-normality for 49,714 out of 184,800 (26.9%) of the time series from cut networks. After correcting for multiple comparisons using the Bonferroni method [46], however, no time series from either the intact or cut networks passed the threshold of p=0.05, suggesting that all the multivariate time series used in our analyses were Gaussian. Data from two nodes in one multivariate time series are plotted as a bivariate histogram in Figure S1 to give a visual of the normality of the data.

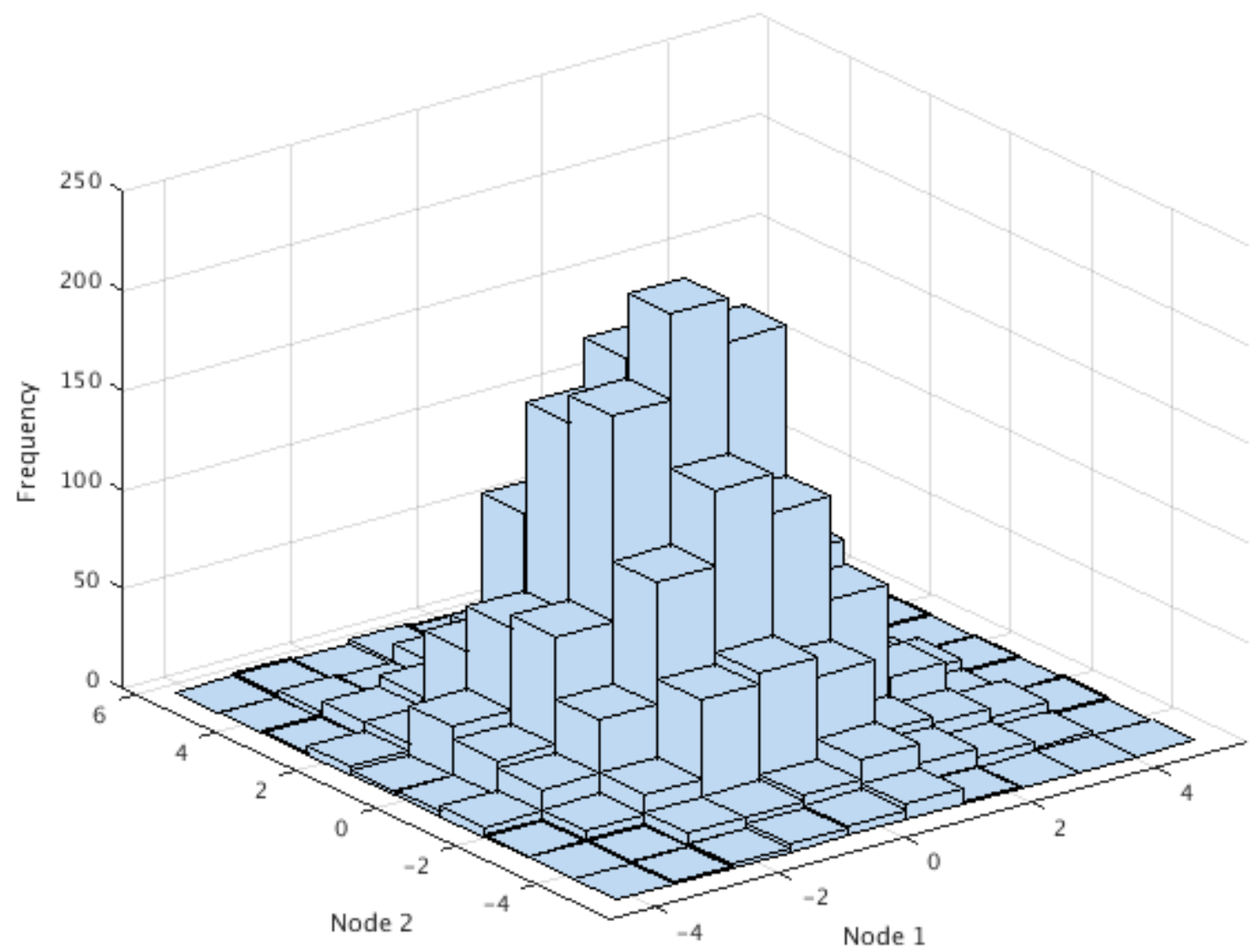

**Figure S1:** Bivariate histogram of the time series of Nodes 1 and 2 from Fig. 4

### S1.4 Network Baseline Integrated-ness Predicts Detection Task Accuracy

Because of the large range of detection task accuracies across networks (Figures 6 and 7), we wanted to know whether there were any network properties that could predict how reliably our measures could detect when a network has been cut. We tested the relationship between detection task accuracy and five network properties: global clustering coefficient, modularity, global efficiency, transitivity, and how integrated a network was at baseline (i.e. its amount of integrated information before being cut). Only baseline integratedness predicted detection task performance (Table S1).

**A**

| CD | | | | |
|---|---|---|---|---|
| | Estimate | Std. Error | t-value | p-value |
| (Intercept) | 149.31 | 170.93 | 0.874 | 0.40509 |
| Mean intact CD | 1032.82 | 290.25 | 3.558 | **0.00613** |
| Global clustering coefficient | -37.62 | 35.61 | -1.056 | 0.31830 |
| Modularity | 44.03 | 56.32 | 0.782 | 0.45440 |
| Global efficiency | -162.57 | 257.45 | -0.631 | 0.54346 |
| Transitivity | 29.18 | 28.25 | 1.033 | 0.32861 |

**B**

| $\phi^{AR}_{MIB}$ | | | | |
|---|---|---|---|---|
| | Estimate | Std. Error | t-value | p-value |
| (Intercept) | 1822.891 | 1262.38 | 1.444 | 0.183 |
| Mean intact $\phi^{AR}_{MIB}$ | 1239.138 | 858.006 | 1.444 | 0.183 |
| Global clustering coefficient | -251.533 | 240.935 | -1.044 | 0.324 |
| Modularity | 671.529 | 390.335 | 01.72 | 0.119 |
| Global efficiency | -2737.011 | 1860.96 | -1.471 | 0.175 |
| Transitivity | -7.408 | 207.793 | -0.036 | 0.972 |

**C**

| $\phi^{AR}_{MMP}$ | | | | |
|---|---|---|---|---|
| | Estimate | Std. Error | t-value | p-value |
| (Intercept) | 953.17 | 1357.52 | 0.702 | 0.5003 |
| Mean intact $\phi^{AR}_{MMP}$ | 3117.52 | 1090.24 | 2.859 | **0.0188** |
| Global clustering coefficient | -295.34 | 285.55 | -1.034 | 0.3280 |
| Modularity | 579.60 | 432.12 | 1.341 | 0.2127 |
| Global efficiency | -1453.67 | 1991.83 | -0.730 | 0.4841 |
| Transitivity | 66.53 | 264.25 | 0.252 | 0.8069 |

**D**

| $\phi^{AR}_{atom}$ | | | | |
|---|---|---|---|---|
| | Estimate | Std. Error | t-value | p-value |
| (Intercept) | 224.93 | 274.14 | 0.82 | 0.433 |
| Mean intact $\phi^{AR}_{atom}$ | 1172.69 | 74.56 | 15.728 | **7.47E-08** |
| Global clustering coefficient | -84.57 | 45.49 | -1.859 | 0.096 |
| Modularity | 113.03 | 82.58 | 1.369 | 0.204 |
| Global efficiency | -375.81 | 403.98 | -0.93 | 0.377 |
| Transitivity | 34.32 | 41.07 | 0.836 | 0.425 |

**E**

| $\phi^{*}_{MIB}$ | | | | |
|---|---|---|---|---|
| | Estimate | Std. Error | t-value | p-value |
| (Intercept) | -208.9 | 249.6 | -0.837 | 0.424 |
| Mean intact $\phi^{*}_{MIB}$ | 10624.31 | 630.75 | 16.844 | **4.11E-08** |
| Global clustering coefficient | -87.13 | 47.63 | -1.829 | 0.101 |
| Modularity | 63.85 | 75.88 | 0.841 | 0.422 |
| Global efficiency | 225.84 | 366.35 | 0.616 | 0.553 |
| Transitivity | 77.09 | 43.22 | 1.784 | 0.108 |

**F**

| $\phi^{*}_{MMP}$ | | | | |
|---|---|---|---|---|
| | Estimate | Std. Error | t-value | p-value |
| (Intercept) | -99.04 | 237.23 | -0.417 | 0.686 |
| Mean intact $\phi^{*}_{MMP}$ | 3169.87 | 237.19 | 13.364 | **3.06E-07** |
| Global clustering coefficient | 19.63 | 50.56 | 0.388 | 0.707 |
| Modularity | 84.81 | 76.59 | 1.107 | 0.297 |
| Global efficiency | 109.12 | 350.25 | 0.312 | 0.762 |
| Transitivity | -72.93 | 43.44 | -1.679 | 0.127 |

**G**

| $\phi^{*}_{atom}$ | | | | |
|---|---|---|---|---|
| | Estimate | Std. Error | t-value | p-value |
| (Intercept) | -53.86 | 241.81 | -0.223 | 0.829 |
| Mean intact $\phi^{*}_{atom}$ | 14556.94 | 873.68 | 16.662 | **4.52E-08** |
| Global clustering coefficient | -57.29 | 43.2 | -1.326 | 0.217 |
| Modularity | 97.82 | 77.09 | 1.269 | 0.236 |
| Global efficiency | 6.4 | 355.07 | 0.018 | 0.986 |
| Transitivity | 12.29 | 39 | 0.315 | 0.76 |

**Table S1:** We performed a multiple linear regression for each measure to see whether its performance on the detection task could be predicted from baseline integrated-ness (i.e., how integrated a measure said a network was when it was intact) or from four graph-theoretic measures: global clustering coefficient, modularity, global efficiency, and transitivity. We found that only baseline integrated-ness explained a significant portion of the variance in accuracy for every measure, excluding $\Phi^{\mathrm{AR}}_{MIB}$, for which none of the variables explained a significant portion of the variance in the accuracy data.